\documentclass{emulateapj}

\def\gtorder{\mathrel{\raise.3ex\hbox{$>$}\mkern-14mu
             \lower0.6ex\hbox{$\sim$}}}

\def\ltsima{$\; \buildrel < \over \sim \;$}
\def\simlt{\lower.5ex\hbox{\ltsima}}
\def\gtsima{$\; \buildrel > \over \sim \;$}
\def\simgt{\lower.5ex\hbox{\gtsima}}

\shorttitle{Black Hole Mass in NGC~6240}
\shortauthors{Medling et al.}

\begin{document}


\title{Mass of the Southern Black Hole in NGC~6240 from Laser Guide Star Adaptive Optics} 


\author{Anne M. Medling\altaffilmark{1,5}, 
S. Mark Ammons\altaffilmark{2,6}, 
Claire E. Max\altaffilmark{1}, 
Richard I. Davies\altaffilmark{3}, 
Hauke Engel\altaffilmark{3}, and 
Gabriela Canalizo\altaffilmark{4}}

\altaffiltext{1}{Department of Astronomy \& Astrophysics, University of California, Santa Cruz, CA 95064, USA; amedling@ucolick.org and max@ucolick.org}

\altaffiltext{2}{Steward Observatory, University of Arizona, Tucson, AZ 85721, USA}

\altaffiltext{3}{Max Planck Institut f{\"u}r extraterrestrische Physik, Postfach 1312, 85741 Garching, Germany; davies@mpe.mpg.de and hauke@mpe.mpg.de}

\altaffiltext{4}{Department of Physics and Astronomy, University of California, Riverside, CA 92521, USA; gabriela.canalizo@ucr.edu}

\altaffiltext{5}{NSF Graduate Research Fellow}
\altaffiltext{6}{Hubble Fellow, Lawrence Fellow}




\begin{abstract}

NGC~6240 is a pair of colliding disk galaxies, each with a black hole in its core.  We have used laser guide star adaptive optics on the Keck II telescope to obtain high-resolution ($\sim 0.06$'') near-infrared integral-field spectra of the region surrounding the supermassive black hole in the south nucleus of this galaxy merger.  We use the K-band CO absorption bandheads to trace stellar kinematics.  We obtain a spatial resolution of about 20 pc and thus directly resolve the sphere of gravitational influence of the massive black hole.  We explore two different methods to measure the black hole mass.  Using a Jeans Axisymmetric Multi-Gaussian mass model, we investigate the limit that a relaxed mass distribution produces all of the measured velocity dispersion, and find an upper limit on the black hole mass at $2.0 \pm 0.2 \times 10^9 M_{\sun}$.  When assuming the young stars whose spectra we observe remain in a thin disk, we compare Keplerian velocity fields to the measured two-dimensional velocity field measured and fit for a mass profile containing a black hole point mass plus a radially-varying spherical component, which suggests a lower limit for the black hole mass of $8.7 \pm 0.3 \times 10^8 M_{\sun}$.    Our measurements of the stellar velocity dispersion place this AGN within the scatter of the $M_{BH}$-$\sigma_{*}$ relation.  As NGC~6240 is a merging system, this may indicate that the relation is preserved during a merger at least until the final coalescence of the two nuclei.  

\end{abstract}


\keywords{Galaxies: kinematics and dynamics -- galaxies: nuclei -- galaxies: interactions -- galaxies: individual (NGC~6240)}


\section{Introduction}

Major mergers are thought to be an important factor in galaxy evolution.  The scenario is as follows: when two gas-rich galaxies of comparable mass collide, large amounts of gas are funneled into the central region, fueling active star formation and nuclear activity \citep[see e.g.][]{SandersMirabel, BarnesHernquist, Genzel, DiMatteo, Hopkins06}.
During this phase of merging, the starbursting galaxy produces copious infrared emission from dust heated by young stars and by the active galactic nucleus (AGN).  We see these galaxies in the local universe as (Ultra-)Luminous InfraRed Galaxies -- (U)LIRGs.  ULIRGs have infrared luminosities of more than $10^{12} L_{\sun}$.  This burst of star-forming activity then uses up much of the gas; the remainder is blown out through a combination of stellar winds and feedback from the AGN.  The scenario then posits that after a major merger, gas-rich galaxies become gas-poor, star formation is largely extinguished, and eventually a ``red and dead" elliptical galaxy is produced with a more massive black hole at its core.

\begin{figure}[h]
\epsscale{0.75}
\plotone{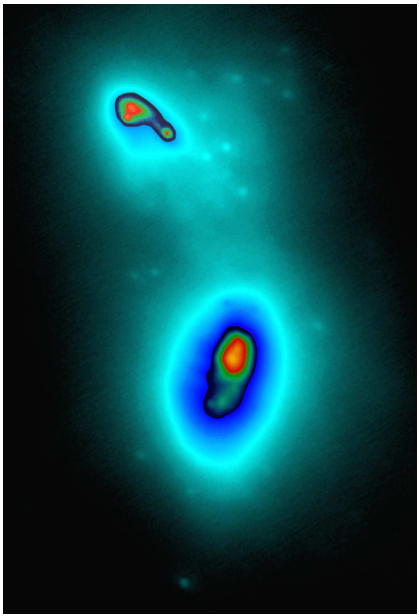}
\caption{Keck adaptive optics image of NGC 6240 in K' band \citep[data first published in][]{Max07}. The dark-blue-enclosed regions in the north and south nuclei, which are separated by about 1.6 arcsec, have each been re-scaled to highlight their interior structure. The more diffuse image of the rest of the galaxy's nuclear region uses a logarithmic color map. Many individual young star clusters can be seen exterior to the two nuclei \citep{Lindsay}. In this image, north is up and east is to the left.}
\label{image}
\end{figure}

NGC~6240 ($z=0.0243$, $d=98$ Mpc for $H_0 = 75$ km s$^{-1}$ Mpc$^{-1}$, 1" = 470 pc), with $L_{IR} \sim 10^{11.8} L_{\sun}$ sits on the boundary between LIRGs and ULIRGs.  Because of its close proximity and spectacular tidal tails and loops, it has become the prototypical example of a gas-rich system in the phase where the two nuclei are close to merging into one.  It has been studied in great detail and in almost every wavelength regime (e.g. x-ray - Komossa et al. 2003; optical - Gerssen et al. 2004; near-IR - Max et al. 2005, 2007; Scoville et al. 2000; Tecza et al. 2000; Engel et al. 2010; mid-IR - Armus et al. 2006; mm - Tacconi et al. 1999; radio - Gallimore \& Beswick 2004, Hagiwara et al. 2011).  Near the core of NGC~6240, the nuclei of the two progenitors are visible 1.2-1.5 arcsec apart, depending on wavelength.  Each of these nuclei holds an AGN; the two sources are resolved in hard x-rays by the Chandra X-Ray Observatory \citep{Komossa}.  The AGNs are deeply obscured at optical wavelengths, however, due to large quantities of dust also present in this region.  By looking into the near-infrared, \citet{Lindsay} have seen young star clusters through some of the dust, products of the most recent close passage of the nuclei (also visible in Figure~\ref{image}).

Supermassive black hole masses are known to scale with certain host galaxy properties, such as bulge light and mass \citep[e.g.][]{KormARAA, KormGeb, Magorrian} and bulge stellar velocity dispersion \citep{Gebhardt, Ferrarese}.  Because these quantities can evolve significantly throughout the major merger process, black hole parameters may also evolve.  It is not known whether the black hole mass grows as a result of the evolution in the host's bulge, or if the bulge growth is moderated by the process of AGN feedback; indeed, it is likely a combination of feeding and feedback processes that maintains these scaling relations.  In order to understand this coevolution, one needs to study systems that are currently merging and eventually to compare these observations to the most detailed merger simulations available.  NGC~6240 represents the ideal candidate for such detailed observations.  

One important aspect of this effort for NGC~6240 is obtaining an accurate measurement of the black hole masses.  A variety of methods are used to measure black hole mass, some of which we review here.  One established method uses full three-integral modeling of stellar orbits; this method, most commonly used in elliptical galaxies, \citep[see e.g.][and references therein]{Siopis, Gult_nuker} uses line-of-sight velocity distributions and light profiles to create a detailed dynamical mass profile of the galaxy, including the black hole mass.  Two- and three-integral models are also sometimes used on bulges of spirals \citep[e.g.][]{Davies06, Onken07, Cappellari09}.  Reverberation mapping \citep[e.g.][and references therein]{Reverb} is a method of measuring black hole masses in AGN; it uses time-resolved brightness fluctuations in the continuum versus lines from the broad-line region to estimate the size of the broad-line region and therefore the mass of the black hole powering the AGN.  Masses measured with reverberation mapping can be calibrated to match the $M_{BH}$-$\sigma_{*}$ relation \citep{Onken04}.  Black hole masses have been derived \citep[][and references therein]{Shields} from the continuum luminosity and the width of the broad H$\beta$ line.  The masses of supermassive black holes powering quasars have been estimated using x-ray luminosity as an indicator, as in \citet{Kiuchi}.  In some cases, it is possible to measure the Keplerian rotation of ionized or molecular gas around a black hole and deduce its mass, as in \citet{Harms}, \citet{HicksMalkan}, and \citet{Neumayer}.  In a few galaxies, masers in the disk allow precise velocity (and therefore black hole mass) measurements \citep{maser, Miyoshi}.  In the Galactic Center, individual stars have been resolved and their orbits around the black hole tracked astrometrically over time to determine the black hole mass \citep{Ghez, Gillessen}.  Recently, adaptive optics have been used to resolve stellar dynamics inside the sphere of influence of black holes in nearby galaxies \citep[e.g.][]{Davies06, Nowak, Nicholas, Gebhardt2011}, a technique which we now build upon.

In a system such as NGC~6240, we are limited in our choice of method; the black holes are obscured by dust and the general system dynamics are unrelaxed because of the ongoing merger.  Keck laser guide star adaptive optics
\citep[LGS AO,][]{Wiz06, vanDam06} enables us to address both of these challenges.  By looking in the near-IR, we look through much of the dust obscuring the relevant kinematics.  The high spatial resolution afforded by the adaptive optics system allows us to focus on stellar dynamics within the sphere of influence of the black hole, unconfounded by the unrelaxed dynamics of the system at large. 


\section{Observations and Data}

We began by observing NGC~6240 with the W.~M. Keck II 10-meter telescope using the Near InfraRed Camera 2 (NIRC2, PIs - K. Matthews \& T. Soifer) and the Keck LGS AO system to obtain high-resolution imaging.  Our images, taken in the K' filter and using the narrow camera (with a 0.01"/pixel plate scale), were previously published in \citet{Max07} and \citet{Lindsay}.  

\begin{figure}[h]
\epsscale{1.2}
\plotone{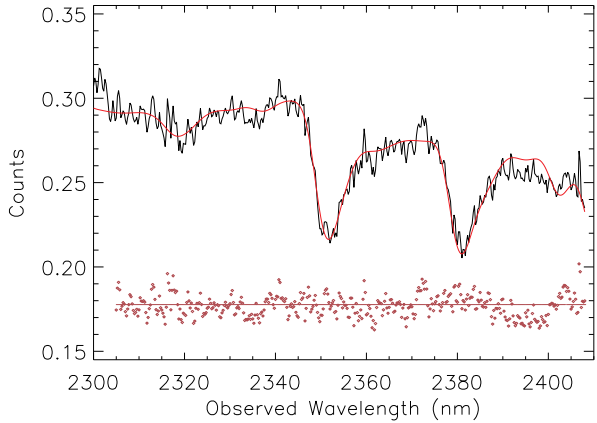}
\caption{An example of the CO absorption bandheads observed in the vicinity of the southern black hole with OSIRIS.  This spectrum was created by binning the light from a 3x3 pixel region around the black hole.  The thick red line overplotted shows the results from template fitting to obtain kinematics, as detailed in Section~\ref{kinematicfitting}.  The red dots along the horizontal line show the residuals of the fit about zero, shifted upwards to fit on the same plot.}
\label{specnearBH}
\end{figure}

We then observed NGC~6240 with the OH-Suppressing InfraRed Imaging Spectrograph \citep[OSIRIS,][]{Larkin06}, on the W.~M. Keck II telescope using LGS AO.  OSIRIS is a near-infrared integral field spectrograph with a lenslet array capable of producing up to 3000 spectra at once.  The spectral resolution ranges from about 3400 in the largest pixel scale to 3800 in the three finer pixel scales; this resolution is sufficient to resolve spectral regions between the OH emission lines from Earth's atmosphere.  Our data are comprised of two 600-second exposures in the Kn5 filter (2.292 $\mu$m - 2.408 $\mu$m) with the 0.035"/pixel plate scale taken on 21 April 2007.  With this filter, we observe the CO (2-0) and (3-1) bandheads at 2.293 $\mu$m and 2.323 $\mu$m rest wavelength (2.345 $\mu$m and 2.380 $\mu$m observed) respectively.  Figure~\ref{specnearBH} shows an example spectrum.  

The Keck LGS AO system uses a pulsed laser tuned to the 589 nm Sodium D$_2$ transition, exciting atoms in the sodium layer of the atmosphere (at $\sim$95 km) and causing spontaneous emission.  Thus, the laser creates a spot in the upper atmosphere which allows the AO system to monitor turbulence below the sodium layer via a Shack-Hartmann wavefront sensor and correct for it with a deformable mirror.  A laser guide star enables high-order corrections to the wavefront, but relies on a natural guide star (which may be fainter and farther away from the target than if no laser were used) to make corrections to image motion (tip and tilt).  Our tip-tilt star (R=13.5 mag) is 35 arcseconds to the northeast of the nuclei.  To estimate the point-spread function (PSF), we took short exposures of our tip-tilt star before and after our observations; these will be described in Section 3.2.

Our OSIRIS data were reduced with the OSIRIS Data Reduction Pipeline v2.2 (available at http://irlab.astro.ucla.edu/osiris), which includes modules to subtract sky frames, adjust channel levels, remove crosstalk, identify glitches, clean cosmic rays, extract a spectrum for each spatial pixel, assemble the spectra into a data cube, correct for atmospheric dispersion, perform telluric corrections, and mosaic frames together.


\section{Methods}

Using the Kn5 filter, we observed the CO (2-0) and (3-1) bandheads at 2.293 $\mu$m and 2.323 $\mu$m rest wavelength respectively.  These molecular features come from the atmospheres of later-type giants and supergiants.  Stellar kinematics are less likely to be disrupted by non-gravitational forces than gas motions, giving us a potentially more robust measurement of the black hole mass.

\subsection{Measuring the Kinematics}
\label{kinematicfitting}

We begin by creating a signal-to-noise (S/N) map using two methods.  We first calculate the S/N theoretically, by adding noise components in quadrature: photon noise from the source, photon noise from the sky, read noise and dark current from characterizations of the detector.  This produces an upper limit to the S/N because other noise sources may be present as well.\\
\begin{equation}\label{SNR}
S/N = {F \gamma  t\over \sqrt{(F  t+ F_{S}  t_{S}  ({t\over t_{S}})^{2})  \gamma + (RN^{2} + ({\gamma\over 2})^{2} + D  t)  n_{pix}}} \\
\end{equation}
In this equation, $F$ is the flux from the galaxy, $F_{S}$ is the flux from the sky, $\gamma$ is the detector gain, $t$ is the exposure time of galaxy frames, $t_{S}$ is the exposure time of sky frames, $RN$ is the readnoise of the detector, $D$ is the dark current, and $n_{pix}$ is the number of spectral pixels.  For a more thorough discussion of this equation, see Section 9.9 in \citet{McLean}.

It is most common to report S/N as an average signal-to-noise ratio \emph{per spectral pixel or per resolution element}.  We adjust the above equation appropriately by dividing the numerator and each variance (\emph{not} the square-rooted noise components) by $n_{specpix}$, the number of spectral pixels included in the region of interest.  In our brightest spatial pixels, we find a S/N per pixel of $\sim40$.  

To confirm our theoretical calculations, we calculate S/N empirically from the spectra.  Bluewards of the CO (2-0) bandhead, we have a spectral region of the galaxy uncontaminated by lines.  Fitting for this continuum level, we find the signal present in each spectrum.  After subtracting off the continuum fit, the root mean square of the residuals gives us a representation of the noise per pixel.  We find good agreement between our empirical and our theoretical estimates of signal-to-noise ratios.

\begin{figure}[h]
\epsscale{1.0}
\plotone{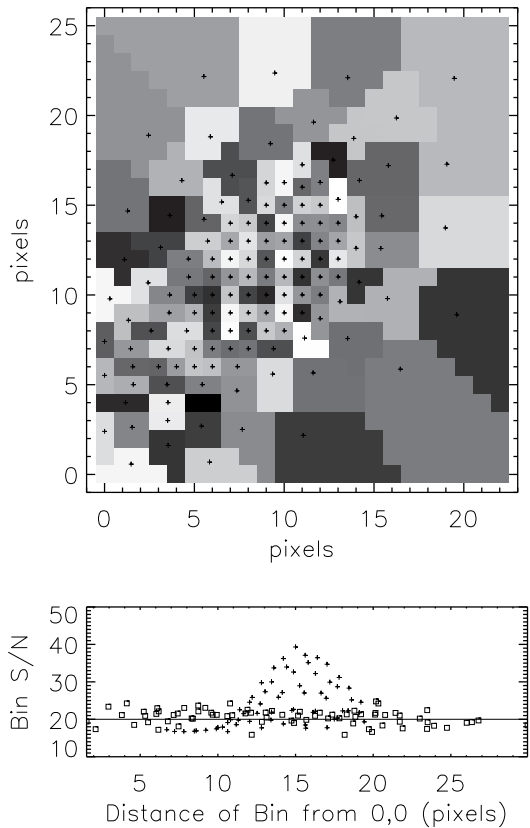}
\caption{Top panel: The Voronoi tesselation bins we impose on our data to equalize to S/N of 20 in the spectrum associated with each spatial bin.  The bins near the center are small (one pixel) because the flux is high enough that the S/N is already above the threshold.  In the outer regions, pixels have been binned together so that the spectra have a S/N of $\sim$20.  Bottom panel: Another representation of the S/N of bins.  Each point represents a bin; the plus symbols are bins that contain only one of the original pixels -- they lie above the threshold line because they had sufficient S/N initially.  The open squares represent bins whose members originally had insufficient S/N; they were binned together until they approximately reach the threshold value.}
\label{voronoi}
\end{figure}

Next we bin our data with optimal Voronoi tesselations using code developed by \citet{Voronoi} to improve the signal-to-noise ratio in our fainter regions.  This algorithm calculates a set of bin centroids according to specific criteria on the topology, morphology, and uniformity of S/N of the final bins.  That is, starting from the unbinned pixel with the highest S/N, the algorithm will include a pixel in the bin if it is adjacent to the starting pixel, does not significantly reduce the ``roundness" of the bin, and, if the bin's S/N is too low, brings the S/N of the final bin closer to the chosen S/N threshold.  In this way, pixels which already have a S/N at or above the chosen threshold are not binned and therefore spatial resolution is not sacrificed unnecessarily.  In lower S/N regions, pixels are binned just enough to provide meaningful measurements.  The morphological requirements create bins that are most likely to share similar velocities and dispersions.  Following \citet{Hauke}, we choose a S/N (per spectral pixel) threshold is 20 for each bin of pixels; the optimal bins are shown in Figure~\ref{voronoi}.

Once we have binned our spectra appropriately, we use the Penalized Pixel Fitting code from \citet{PPXF} to fit radial velocities to each spectrum.  This method implements a maximum penalized likelihood approach for extracting stellar kinematics from absorption-line spectra.  The algorithm parametrically expands the line-of-sight velocity distribution as a Gauss-Hermite series and allows for the choice of a penalty against higher-order moments.  This penalty will bias the fit against higher-order moments, so that the fit must be improved by a certain specified amount to include them.  In this way, higher-order fits are possible where spectra have a high enough signal-to-noise ratio, but the fits will tend to simple Gaussians in the low signal-to-noise limit.  In order to trust higher-order moments, a signal-to-noise ratio of $>50$ \citep{Cappellari09} is usually required.  Since binning our spectra up to a S/N of 50 would decrease our spatial resolution signficantly, we choose to fit simple Gaussians.

\begin{figure}[h]
\epsscale{1.1}
\plotone{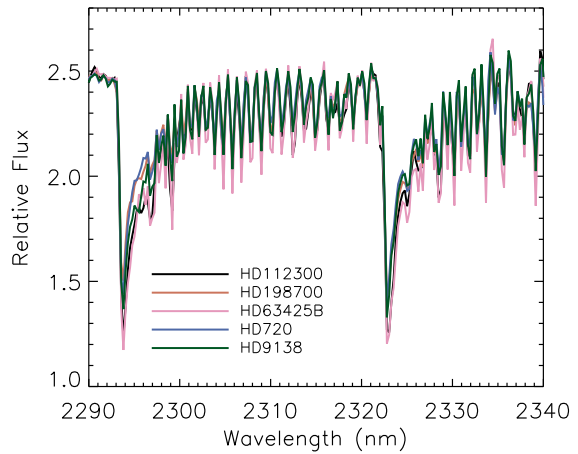}
\caption{Spectra of the five late-type giants and supergiants used as stellar velocity templates in our dynamical modeling.  The deep CO bandheads match those seen in our spectra (Figure~\ref{specnearBH}).  The templates were shifted in wavelength to correct for peculiar velocities of the stars, to match the CO transitions in a vacuum.}
\label{templates}
\end{figure}

One key feature of this code is the option of including a set of stellar templates from which to fit the kinematics.  There has been some debate on which templates most accurately represent the stellar populations of NGC~6240.  \citet{Tecza} conclude that the bulk of the light in the near-infrared is due to late K or early M supergiants, while \citet{Hauke} argue that late-type giants characterize the global stellar population better.  With Cappellari's method, we are able to input a variety of stellar templates and allow the parametric fit to select the best combination of templates for each spectrum.  In our final iteration of the code, we selected 5 stellar templates from the GNIRS library \citep{GNIRS} that exhibited the deepest CO bandheads.  The stellar templates chosen were: HD112300 (spectral type M3III), HD198700 (K1II), HD63425B (K7III), HD720, (K5III), and HD9138 (K4III).  The region of these spectra around the CO bandheads is shown in Fig~\ref{templates}.  

\begin{figure}[h]
\epsscale{1.2}
\plotone{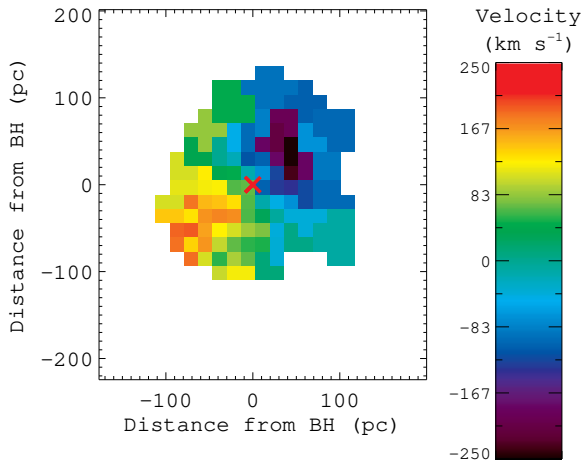}
\caption{Stellar velocity field measured from the OSIRIS IFU data near the southern black hole (red x).  Typical errors are $\sim 10$ km s$^{-1}$ in the brightest regions and $\sim 30$ km s$^{-1}$ overall.  Large pixels on the periphery have been binned using Voronoi tesselation to improve S/N (see Figure~\ref{voronoi}).  Voronoi bins with centroids further than $\sim 115$ pc from the black hole (in projection) have been colored white to mask out regions which are less affected by the black hole's gravity.  (The data used to create this figure are available for download in the online journal.)}
\label{vel}
\end{figure}

\begin{figure}[h]
\epsscale{1.2}
\plotone{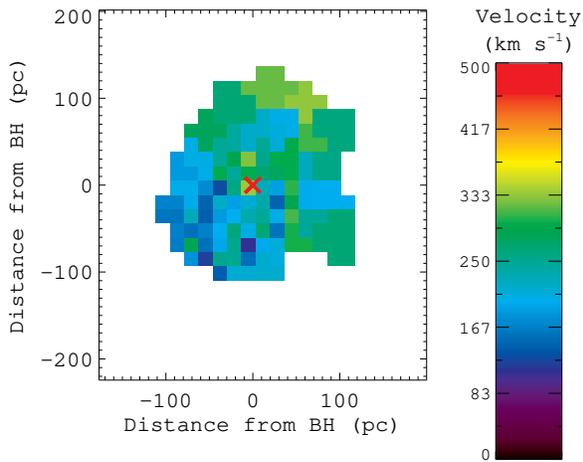}
\caption{Stellar velocity dispersion measured from the OSIRIS IFU data near the southern black hole (red x).  Typical errors are $\sim 10$ km s$^{-1}$ in the brightest regions and $\sim 70$ km s$^{-1}$ overall.  Large pixels on the periphery have been binned using Voronoi tesselation to improve S/N (see Figure~\ref{voronoi}).  Regions far from the black hole have been masked out, as in Figure~\ref{vel}.  (The data used to create this figure are available for download in the online journal.)}
\label{disp}
\end{figure}

We show the measured velocity and velocity dispersion maps from OSIRIS in Figures~\ref{vel} and \ref{disp}, respectively. We estimate the errors in the velocity and velocity dispersion measurements by running a Monte Carlo simulation, adding the appropriate amount of random noise to each spectrum and refitting them 100 times.  Formal measurement errors in our velocity and velocity dispersion measurements are $\sim 10$ km s$^{-1}$ in the brightest regions, and average about 30 and 70 km s$^{-1}$, respectively, over the whole fitting region.

With velocity and velocity dispersion maps, it is tempting to think about the measured $v/\sigma $ of the system, to determine how much of the stellar dynamical energy is in rotation.  However, this must be handled carefully in a system like NGC~6240.  In our data, we measure velocity peaks of $\sim 200$ km s$^{-1}$, and velocity dispersion that varies between roughly 200 and 300 km s$^{-1}$, suggesting that the energy is approximately evenly divided between ordered rotation and random orbits, with perhaps up to 50\% more in the latter.  However, in this region of NGC~6240, we know there are clumps of intervening material such as spiral arm remnants or tidal tails \citep[e.g.][]{Hauke}, which would cause an increase in the measured velocity dispersion that is not yet inherent to either the nuclear disk or the spheroid.  Because we do not have a way of determining how much of the velocity dispersion measured is from relaxed material, our analysis yields only a lower bound on $v/\sigma $.  We conclude that a significant fraction of the kinetic energy is in rotation.

Once we have maps of velocity and velocity dispersion, we compare our data to models that contain a black hole.  Each model, of course, comes with its own set of assumptions, and in a late-stage merging system such as NGC~6240, we must think carefully about what such models can tell us.  While the quality of these data is clear, understanding how to analyze them is not straightforward. Here we discuss two possible ways to measure the black hole mass from these data and compare the results of the two methods.

\subsection{Dynamical Analysis - JAM Modeling}

We begin our analysis by utilizing the JAM modeling code \citep[Jeans Anisotropic Multi-Gaussian expansion dynamical models][]{JAM}, a technique based on the two-integral axisymmetric Jeans formalism but which has been expanded to allow for anisotropy via the parameter $\beta _{z} = 1 - (\sigma _{z} / \sigma _{R} )^{2}$.  JAM modeling efficiently utilizes the axisymmetric dynamics seen near the south nucleus, and does not require higher-order Hermite moments.

This method, which fits $v_{rms} = \sqrt{v^{2} + \sigma ^{2}}$, is likely to overestimate the black hole mass by assuming the dynamics measured belong to a relaxed system.  In fact, an unknown fraction of the measured velocity dispersion is due to intervening material, such as tidal tails, that has not yet reached dynamical equilibrium.  Additionally, this method assumes axisymmetric and smooth light and mass profiles.  Finally, we assume a constant mass-to-light ratio.

The JAM modeling code requires a high-resolution light profile, which we parametrize using the Multi-Gaussian Expansion code \citep[MGE][]{MGE} designed to work with the JAM code.  To fit our light profile over a larger field of view than is available in our OSIRIS data, we use our K' NIRC2 imaging and mirror it about the minor axis of the nucleus.  We do this because the southeast side of the nucleus is considerably less extincted, so our signal-to-noise ratio is much improved.  Once we have symmetrized the observed light profile, we then de-extinct this using the extinction map of Figure 9 in \citet{Hauke}.  Using this extinction map, our recovered intrinsic brightness peaks at the same location as the kinematic center of our dynamical data, which means that in K-band the nucleus is only partially extincted.  (In contrast, the extinction at visible wavelengths is so severe that the entire region surrounding the south black hole cannot even be seen \citep{Max05}.)  This is important, as most of the orbital information reported by JAM modeling is contained in the light profile.  In order to determine the appropriate mass-to-light ratio, anisotropy parameter $\beta _{z}$ and black hole mass, we compare the resulting dynamical models to those measured from our OSIRIS data in K-band.

\begin{figure}[h]
\epsscale{1.2}
\plotone{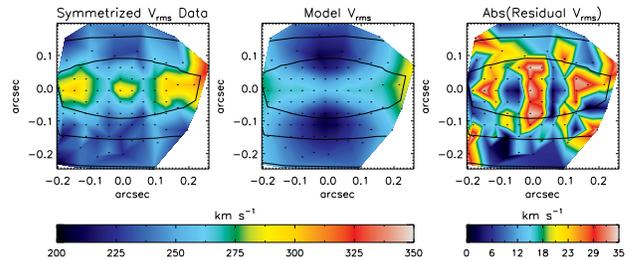}
\caption{Left: Symmetrized $v_{rms}$ map \citep[as in][]{JAM} from OSIRIS observations.  Center: Best-fitting axisymmetric JAM model, on the same scale and color bar as data (left).  Right: Map of residuals (absolute value of difference between left and center panels) with new color bar to the (right).  In all panels, the black contours show the Gaussian expansion of the light profile, and the black dots represent the centroid of each spatially-binned spectrum, where velocity and velocity dispersion are measured.  Axes show distance from the central black hole in arcseconds.}
\label{JAM}
\end{figure}

Our best-fit models for this method measure a black hole mass of $2.0 \pm 0.2 \times 10^9 M_{\sun}$, and are shown, along with the symmetrized $v_{rms}$ data for comparison, in Figure~\ref{JAM}.  The reduced $\chi^{2}$ statistic, fitting over 120 points, is 2.21.

\subsection{Dynamical Analysis - Thin Disk}

As a sanity check on the JAM model, and to provide a lower-limit to the black hole mass, we explore a simple model comparing the velocity field to that of a thin disk with a given enclosed mass profile exhibiting Keplerian rotation.  

In this method we do not include a dispersion component because intervening material may inflate that measurement.  Here we assume that the energy in intrinsic dispersion in the nuclear stellar disk is negligible compared to the energy in rotation.  This should give a lower limit to the black hole mass.

We are measuring the dynamics of young stars in the very nucleus of a gas-rich merger.  A thin disk of young stars may be expected because they can form out of the nuclear disks of gas and dust seen in merger simulations by, for example, \citet{Mayer, KazMayer, HQ2010} and in observations by \citet{gasdisk, M31}.  We clearly see in the OSIRIS velocity field a sharp steepening of the velocity gradient in the region of the black hole: a sign of strong rotation, and reminiscent of a thin Keplerian disk embedded in a larger spherical potential.   

We focus our mass modeling on the region of the sphere of influence of the black hole, where the dynamics are most likely to be well-behaved.  We begin with a thin Keplerian disk model, using $v = \sqrt{{G M_{BH}}/{r}}$.  This will attribute all mass enclosed to the black hole, and therefore overestimate the mass; we only use this case to check the sub-pixel position of the black hole and get a rough set of parameters over which to fit a more complex model, which includes a spheroid, described in Section~\ref{parameters}.

Once a model velocity field is constructed for a specific set of parameters, we use that information to create a synthetic datacube: for each spaxel, we use a template CO bandhead spectrum shifted in velocity space to the appropriate velocity and weighted by the total flux in that spaxel.  With this synthetic datacube we perform a wavelength-by-wavelength convolution with the PSF to simulate the residual smearing from the atmosphere and optical system.  Convolving each wavelength slice with the PSF models the true image blurring, as opposed to smoothing the overall velocity field.  Once the datacube has been convolved in this way, we remeasure the velocities from each spatial pixel's spectrum to derive the smoothed velocity field.  We compare the resulting velocity field with the observed velocities measured by OSIRIS.  We repeat this process, making models for various parameter sets, and compare each with the observed data.

\subsubsection{Model Parameters}
\label{parameters}

At these spatial scales, the velocity field of a thin Keplerian disk around the black hole can be fully described by a few parameters: black hole position, disk inclination, disk position angle, and enclosed mass.  We fit the position angle of the velocity field first, as it may be fit largely independently of the other black hole parameters.  Our tests indicate that the positive velocity peak falls at a PA of $130^{\circ}$ measured counter-clockwise from north, relative to the black hole's position.  We adopt this position angle in our later fits.  The position of the black hole is not trivial to pinpoint.  Because of 6-8 magnitudes of extinction distributed unevenly across the nucleus \citep{Hauke}, we avoid using our K'-band isophotes to find the center of the southern nucleus, and instead use the measured velocities to determine the kinematic center.  We have a close estimate of the position of the southern black hole using relative astrometry from the northern black hole, whose position is visible in our data \citep{Max07}.    To improve our positional accuracy, we allow the black hole position to vary in the fit by sub-pixel amounts in a simple code that only takes into account the black hole mass.  We then fix the position of the black hole, along with our PA, for our full suite of models.

\begin{figure}[h]
\epsscale{1.0}
\plotone{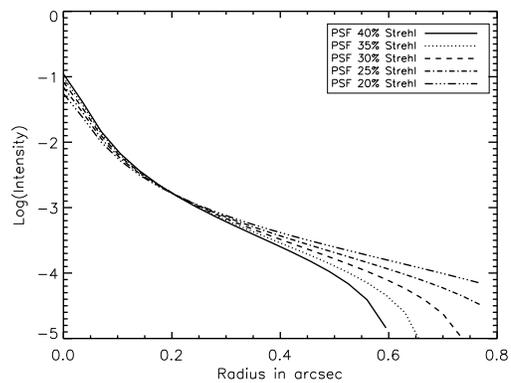}
\caption{The five PSFs used by our model fitting routine.  The Strehl ratio is varied as a free parameter in order to avoid systematic errors in black hole mass measurement based on PSF mismatch, while maintaining a FWHM of 65 mas, to match the FWHM measured in the data.}
\label{PSFs}
\end{figure}

As AO measurements can be affected by the PSF achieved in our observations, we must carefully consider the PSF used in our models.  We use short exposures of our tip-tilt star bracketing our observations which are useful to characterize the performance of the AO system and the conditions during the evening.  Our first PSF model is a Moffat fit of the tip-tilt star, which has a Strehl ratio of 20\% and FWHM of 65 mas.  We find that the north nucleus point source has a FWHM of 63 mas, consistent with that of the tip-tilt star.  However, the tip-tilt star gives a lower-limit to the actual Strehl ratio since, for shorter exposure times appropriate to the tip-tilt star, the low-bandwidth wavefront sensor \citep{Wiz06} did not have time to settle into the most accurate correction.  The longer exposures on NGC~6240 itself do provide sufficient settling time.  This effect is partly offset by additional blurring due to the offset between the tip-tilt star and NGC~6240 (anisokineticism).  Because the distance between the tip-tilt star and the nuclei is only 35", this effect is expected to be modest \citep{vanDam06}, and does not completely counteract the aforementioned effect from the low-bandwidth wavefront sensor.  We see that the AO system performs better during our longer galaxy exposures.  To accommodate this, we adjust the Moffat fit coeffecients to Strehls of 25\%, 30\%, 35\% and 40\%, while maintaining the FWHM (see radial profiles in Figure~\ref{PSFs}).  The Strehl ratio then becomes a parameter in our model-fitting procedure, in which we create models with each of these PSFs to compare with our data.

In an asymmetric, dusty system such as NGC~6240, we must also carefully consider the region over which we compare our models to the data, masking out regions that are less important to match.  We expect that our model will most accurately describe the velocities close to the black hole.  At some distance from the black hole, we expect other galaxy and tidal components to dominate, which may affect the model fit.  To test this concern, we use a variety of different masks to vary the distance from the black hole at which we stop our comparison.  We find that the best-fitting enclosed mass is not sensitive to the mask selection.

Our mass profile includes a radially-varying contribution to the mass profile, representing a spheroidal component of the galaxy.  We construct models that include both a point mass at the center and a spherically-symmetric mass profile to mimic the inner regions of the bulge.  We parametrize the spheroidal component as $\rho(r) = \rho_{0} r^\gamma$, where $\rho$ is the mass density.  We fit three parameters to the enclosed mass profile: the black hole mass $M_{BH}$, the normalization for the spheroidal compontent $\rho_{0}$, and the power-law index of the spheroidal component, $\gamma$.

\begin{figure}[h]
\epsscale{1.1}
\plotone{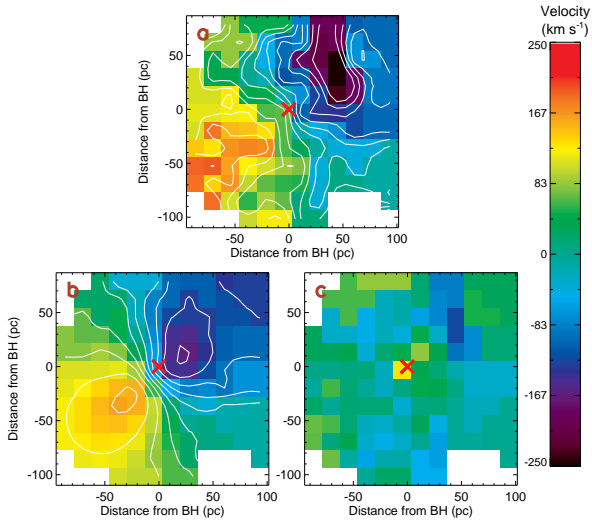}
\caption{(a) The observed velocity field of our OSIRIS data for comparison with our best-fit models. (b) The best-fitting velocity field for a mass model containing a point mass and an extended, spherically-symmetric component.  (c) The residuals found by subtracting the observed velocity field from the model shown in (b).  Note that the model yields a good fit to the central steep velocity gradient associated with the black hole.  In each panel regions far from the black hole have been masked out, as in Figure~\ref{vel}.}
\label{velobsmodel}
\end{figure}

\subsubsection{Results of Thin Disk Model Fitting}

Our best fit gives a black hole mass of $8.7 \times 10^8 M_{\sun}$ with a spheroid of $2.9 \times 10^8 M_{\sun}$ within 100 pc of the black hole and an index of $\gamma = 1.5$ (see velocity field of this model in Figure~\ref{velobsmodel}).  The reduced $\chi^{2}$ statistic for this model is 3.7, fitting over 65 velocity datapoints.  To estimate the accuracy of this measurement, we perform a Monte Carlo simulation, fitting models to our observations with 100 different representations of noise added.  The distribution of fitted masses can be well-fit by a Gaussian, the width of which represents our one-sigma error bars.  Including this, our best fit shows a system with a black hole mass $8.7 \pm 0.3 \times 10^8 M_{\sun}$.

\subsubsection{Trends in the Reduced $\chi^{2}$ Map}

\begin{figure}[h]
\plotone{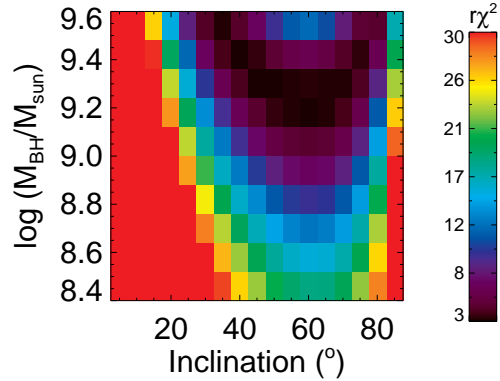}
\caption{The map of reduced $\chi^{2}$ statistics for our thin disk model as the black hole mass (vertical axis) and the inclination (horizontal axis) are varied.  The remaining parameters (spheroid paramaters $\rho_0$ and $\gamma$, velocity offset, and PSF) are held fixed.  We see that as the inclination decreases (the disk of stars becomes more face-on), a larger black hole mass is required to match the observed radial velocity peaks for inclinations $\lesssim 60^{\circ}$.  As the inclination increases beyond $60^{\circ}$ (the disk of stars becomes more edge-on), we also see the best-fit black hole mass increase; this is likely because a larger black hole is required to affect off-axis regions of the spider diagram when the disk is more edge-on.  These fits were measured with 65 datapoints.}
\label{chi2imbh}
\end{figure}

It is also instructive to look at trends in the reduced $\chi^{2}$ maps.  The simplest example is shown in Figure~\ref{chi2imbh}, demonstrating that, for a thin disk model such as this one, the black hole must be more massive to match the velocity field when the inclination is lower, for $i < 60^{\circ}$.  This map also shows that the black hole mass increases again at the highest inclinations, as the model tries to match the velocity field not only along the major axis but in the surrounding spatial pixels as well (the well-known ``spider diagram").  If NGC~6240 were dust-free, the morphology of the region around the south nucleus could, in principle, be used to independently measure (or constrain) the inclination.  However, the northwest half of the circumnuclear disk is heavily obscured in NGC~6240.  In this case, the image can only provide a lower-limit for the inclination ($\sim55^\circ$).  

\begin{figure}[h]
\plotone{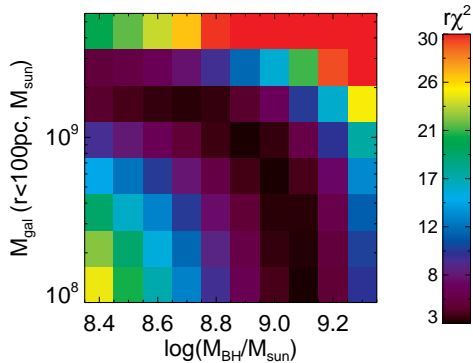}
\caption{The map of reduced $\chi^{2}$ statistics as the black hole mass component (horizontal axis) and the spherically-symmetric mass component (vertical axis) are varied.  The other parameters (inclination, density profile index $\gamma$, and velocity offset) are held fixed.  We see that on the right side (at high black hole mass), the best fitting models have less mass in the spheroidal component.  On the left side (with black hole mass becoming negligible), the mass of the spheroidal component flattens out at $2 \times 10^9 M_{\sun}$, the total enclosed mass at our resolution limit.  These fits were measured with 65 datapoints.}
\label{chi2mm}
\end{figure}

We see a demonstration of the trade-off of mass between the spheroid and black hole by looking at a 2-dimensional map of reduced $\chi^{2}$ statistics (Figure~\ref{chi2mm}), varying the amount of mass that goes into each component of our mass profile.  As more mass is put in the black hole, models with less mass in the radial component fit better; as less mass is put in the black hole, the mass in the radial component must increase to best fit the data.  In each case, the combined mass reaches a maximum at $\sim 2 \times 10^9 M_{\sun}$ within 100 pc of the black hole. 

Lastly, we consider our choice of varying PSFs in our model fitting.  Our best-fit models from this technique prefer PSFs with a Strehl ratio of 25\%, near the lower limit of what we would expect from our data.  We compare these to models with 40\% Strehl, approximately the best PSF we could hope for under the conditions of these observations.  When forcing an improved PSF, the best-fitting black hole mass increases slightly, to $9.2 \times 10^8 M_{\sun}$.  This is only slightly outside of the 1-$\sigma$ range, and has a reduced $\chi^{2}$ statistic of 4.8, considerably higher than our better-fitting models.  While we are confident that fitting for PSF was a good choice, we also note that the black hole mass is not very sensitive to this parameter.


\section{Discussion}

\subsection{How Much Mass Could Be Due to a Nuclear Star Cluster?}

It is important to consider the limitations of our approaches.  With a black-hole-only thin disk model, one does not directly measure the black hole mass; one measures the mass enclosed in the central OSIRIS pixel, 17 parsecs on a side.  Including a spheroidal component implicitly assumes that the stellar mass density profile is smooth and that all other mass is due to the black hole.  What fraction of this mass might be due instead to a nuclear star cluster?

To address this concern, we refer to our high-resolution NIRC2 K' imaging.  Through a careful deconvolution by the PSF, we arrived at a cleaned image of the south nucleus.  We flux-calibrated this image by matching the large-scale luminosity to that reported in \citet{Hauke}, and assumed their mass-to-light ratio of 1.9.  

This analysis showed that there could be up to $3 \times 10^8 M_{Sun}$ of stellar mass within the sphere of influence of the black hole, consistent with the mass in our fitted spheroidal component.  

\subsection{The $M_{BH} - \sigma_{*}$ Relation}

With our measurement of the black hole mass, it is interesting to consider where NGC~6240 would fall on the $M_{BH} - \sigma_{*}$ relation, which compares the black hole mass to the stellar velocity dispersion of the bulge of the host galaxy.  Because NGC~6240 is a merging system, it plausibly lacks a relaxed bulge component; therefore it is not easy to define exactly where or how the velocity dispersion should be measured to compare to the $M_{BH} - \sigma_{*}$ relation.  Still, it is worth looking at merging active systems since they are in the process of evolving along these relations.  Here we are able to take our first glimpse of where a system might fall on these relations while in the process of merging.  Does the black hole grow more quickly than the larger-scale galactic properties, or must it play catch-up after the galaxy's bulge has settled back to an equilibrium state?

The $M_{BH} - \sigma_{*}$ relation is well-defined only in systems that are dynamically-relaxed on the large scale.  In such a system, the central black hole mass can be compared to the integrated velocity dispersion of stars in the bulge within one effective radius.  NGC~6240 has two black holes; here we only consider the southern one.  It is not trivial to measure the equivalent stellar velocity dispersion associated only with this black hole in the southern galaxy, as the bulges of both progenitor galaxies have begun to merge.  We expect the stellar velocity dispersion at this stage to be low compared to the final value; quite a bit of energy is still in ordered rotation and has yet to be randomized.  As reported in \citet{Hauke}, the south nucleus shows a maximum rotational velocity of $\sim 300$ km s$^{-1}$, putting as much or more energy in ordered rotation as in dispersion.  Our spectra also give us only the dynamics of later-type giants and supergiants, via the CO bandheads.  \citet{Rothberg} have suggested that this also gives an underestimate of velocity dispersion as compared to measurements made using the Ca triplet absorption lines at 8500 \AA.  

Our spectra also only give us dynamics very close to the nucleus.  We bin our OSIRIS data into one spectrum encompassing our entire south nucleus (inside a radius of $\sim 300$ pc); the measured velocity dispersion from the CO bandheads is $282 \pm 20$ km s$^{-1}$.  For comparison, we make the same measurement using SINFONI data from \citet{Hauke}; this velocity dispersion, $310 \pm 12$ km s$^{-1}$, encompasses a wider region ($r \leq 500$ pc).  We also compare these two values of stellar velocity dispersion to previous measurements from the literature.  \citet{Tecza} measured $\sigma_{*}\sim$ 236 km s$^{-1}$ using the CO bandheads within 235 pc of the south nucleus.  \citet{OOMM99} measured the stellar velocity dispersion integrated over the entire system, but used three different absorption lines: Si 1.59 $\mu$m (313 km s$^{-1}$), CO 1.62 $\mu$m (298 km s$^{-1}$), and CO 2.29 $\mu$m (288 km s$^{-1}$).  Estimates of $\sigma_{*}$ made from integrating over a larger fraction of the galaxy could be higher because they combine multiple dynamical populations (e.g. the north and south nuclei and intervening spiral arms); however, such estimates do not distinguish between material associated with this black hole and the northern black hole, as both are partially within the seeing-limited PSF.

\begin{figure}[h]
\epsscale{1.1}
\plotone{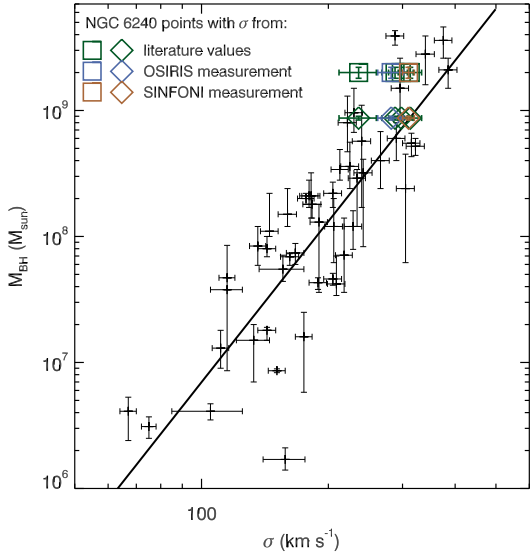}
\caption{Plot of the $M_{BH} - \sigma_{*}$ relation as recently recalculated by \citet{Gult}.  The small black points and error bars represent the black hole mass measurements from the literature.  The larger colored squares and diamonds represent our black hole mass measurements in the south nucleus of NGC~6240, using different values of $\sigma_{*}$.  Squares show the black hole mass measurement from JAM modeling, and represent an upper limit to the black hole mass.  Diamonds show the black hole mass measurement from our thin disk approximation, and represent a lower limit to the black hole mass.  The dark green points are our mass measurement paired with measurements of $\sigma_{*}$ from \citet{Tecza} and \citet{OOMM99}.  The blue and brown points plot our measurements of $\sigma_{*}$ from OSIRIS and SINFONI respectively.  See text for further details.}
\label{Msigma}
\end{figure}

We plot our black hole mass measurement on the $M_{BH} - \sigma_{*}$ relation along with other dynamical black hole mass measurements compiled by \citet{Gult} in Figure~\ref{Msigma}.  We plot separate points for our black hole mass measurements under two different assumptions.  Because there is some ambiguity about the appropriate way to measure the stellar velocity dispersion in the bulge in a system such as NGC~6240, we plot our measured black hole mass with several different $\sigma_{*}$ values from the literature as well as with our measured $\sigma_{*}$ very close to the black hole.

NGC~6240 appears to lie well within the scatter of the $M_{BH} - \sigma_{*}$ relation, which may suggest that the black hole mass and the bulge velocity dispersion grow simultaneously and at similar rates during a major merger.  NGC~6240 is a late-stage merger, but these data could also suggest that a system doesn't evolve along the $M_{BH} - \sigma_{*}$ relation until the very end stages of a merger, perhaps during nuclear coalescense.  To test this, more systems at this and later stages of merging will have to be studied.  This is interesting to compare to galaxy merger simulations; for example, \citet{Dasyra06} conclude that, if the accretion efficiency stays constant, the $M_{BH} - \sigma_{*}$ relation should be maintained from midway between the first encounter and coalescence through to final relaxation.  New higher-resolution simulations are being performed by Stickley et al. (in prep) which show that the stellar velocity dispersion stays low in the nuclear regions much longer than in the rest of the system; results will follow young and old stellar populations separately to help understand the dynamics measured from specific lines.


\section{Conclusions}

Black holes are an important ingredient in galaxy evolution, as seen by the tight correlations between black hole mass and host galaxy properties.  Hydrodynamic simulations of galaxy mergers have suggested that gas-rich mergers, which can drive galaxy evolution, can provide fuel for black hole accretion.  In turn, these accreting black holes can radiate enough energy to affect the surrounding galaxy.  In light of these discussions, it is particularly interesting to study systems that appear to be in the middle of such an evolutionary event.  NGC~6240 represents a nearby galaxy such as this: a merging gas-rich system with two actively accreting black holes.  However, such systems are notoriously difficult to study because of their unrelaxed dynamics and their dusty cores.

We have presented high-spatial resolution kinematics within the sphere of influence of the black hole in the south nucleus of NGC~6240, a nearby late-stage merger, made possible by laser guide star adaptive optics.  For this test case, we have explored two possible methods for measuring black hole mass in such a system and compared their results and assumptions.

We have utilized the JAM modeling technique made public by \citet{JAM}, demonstrating that it is possible to complete such an analysis on a late-stage merger.  We point out that this technique likely overestimates the black hole mass by assuming that all measured velocity dispersion is due to a relaxed system (ignoring intervening unrelaxed material, e.g. tidal tails), and therefore report an upper limit of $2.0 \pm 0.2 \times 10^9 M_{\sun}$.  To provide a lower limit to the black hole mass, we explore the opposite assumption: that all velocity dispersion is caused by intervening material, and that the young stars sit in a thin disk around the black hole.  This model of Keplerian rotation around a black hole plus smooth spheroidal mass profile suggests a black hole which is at least $8.7 \pm 0.3 \times 10^8 M_{\sun}$.

We find that these two techniques provide measurements that are roughly consistent, and that follow the biases implied by their intrinsic assumptions.  To determine which set of assumptions is more reliable would require a detailed study of high-resolution galaxy simulations, which is beyond the scope of this paper.  Still we are encouraged that, to within a factor of about two, both measurements agree.

While we cannot make generalizations on how all black hole-galaxy coevolution must proceed, it seems that in this case, the black hole and the host galaxy parameters grow together along the $M_{BH} - \sigma_{*}$ scaling relation, instead of one preceeding the other.  We are beginning an observing campaign to study other local merging galaxies using the same techniques, to investigate a larger sample size.  
  
  
\acknowledgements
We enthusiastically thank the staff of the W. M. Keck Observatory and its AO team, for their dedication and hard work.  Data presented herein were obtained at the W. M. Keck Observatory,
which is operated as a scientific partnership among the California Institute of Technology, the University of California, and the National Aeronautics and Space Administration.  The Observatory and the Keck II Laser Guide Star AO system were both made possible by the generous financial support of the W. M. Keck Foundation.  The authors wish to extend special thanks to those of Hawaiian ancestry on whose sacred mountain we are privileged to be guests.  Without their generous hospitality, the observations presented herein would not have been possible.  This work was supported in part by the National Science Foundation Science and Technology Center for Adaptive Optics, managed
by the University of California at Santa Cruz under cooperative agreement AST 98-76783.  This material is based in part upon work supported by the National Science Foundation under award number AST-0908796.  We thank the referee for very helpful insights.  AM would also like to thank Aaron Romanowsky and Nicholas McConnell for helpful discussions.  AM is supported by a Graduate Research Fellowship from the National Science Foundation.  GC acknowledges support from NSF grant AST 0507450.  SMA acknowledges support from Program number HST-HF-51250.01-A, provided by NASA through a Hubble Fellowship grant from the Space Telescope Science Institute, which is operated by the Association of Universities for Research in Astronomy, Incorporated, under NASA contract NAS5-26555.

{\it Facility:} \facility{Keck:II (Laser Guide Star Adaptive Optics, OSIRIS)}


\begin{thebibliography}{}

\bibitem[Armus et al.(2006)]{Armus} Armus, L., et al.\ 2006, 
\apj, 640, 204 

\bibitem[Barnes \& Hernquist(1996)]{BarnesHernquist} 
Barnes, J.~E., \& Hernquist, L.\ 1996, \apj, 471, 115

\bibitem[Binney 
\& Tremaine(1987)]{BT} Binney, J., \& Tremaine, S.\ 1987, Galactic Dynamics (Princeton, NJ: Princeton University Press)

\bibitem[Bonnet et al.(2004)]{SINFAO} Bonnet, H., et al.\ 
2004, \procspie, 5490, 130 

\bibitem[Cappellari(2002)]{MGE} Cappellari, M.\ 2002, 
\mnras, 333, 400 

\bibitem[Cappellari 
\& Copin(2003)]{Voronoi} Cappellari, M., \& Copin, Y.\ 2003, \mnras, 342, 345 

\bibitem[Cappellari 
\& Emsellem(2004)]{PPXF} Cappellari, M., \& Emsellem, E.\ 2004, \pasp, 116, 138 

\bibitem[Cappellari(2008)]{JAM} Cappellari, M.\ 2008, 
\mnras, 390, 71 

\bibitem[Cappellari et al.(2009)]{Cappellari09} Cappellari, M., 
Neumayer, N., Reunanen, J., van der Werf, P.~P., de Zeeuw, P.~T., 
\& Rix, H.-W.\ 2009, \mnras, 394, 660 

\bibitem[Dasyra et al.(2006)]{Dasyra06} Dasyra, K.~M., et al.\ 
2006, \apj, 651, 835 

\bibitem[Davies(2008)]{Davies08} Davies, R.\ 2008, New Astronomy 
Review, 52, 307 

\bibitem[Davies et al.(2006)]{Davies06} Davies, R.~I., et al.\ 
2006, \apj, 646, 754 

\bibitem[Denney et al.(2009)]{Reverb} Denney, K.~D., et al.\ 
2009, \apj, 702, 1353 

\bibitem[Di Matteo et al.(2005)]{DiMatteo} Di Matteo, T., 
Springel, V., \& Hernquist, L.\ 2005, \nat, 433, 604 

\bibitem[Eisenhauer et al.(2003)]{SINF} Eisenhauer, F., et 
al.\ 2003, \procspie, 4841, 1548 

\bibitem[Engel et 
al.(2010)]{Hauke} Engel, H., et al.\ 2010, \aap, 524, A56

\bibitem[Ferrarese \& Ford(2005)]{FF05} Ferrarese, L., 
\& Ford, H.\ 2005, Space Science Reviews, 116, 523

\bibitem[Ferrarese \& Merritt(2000)]{Ferrarese} Ferrarese, L., 
\& Merritt, D.\ 2000, \apjl, 539, L9

\bibitem[Gallimore \& Beswick(2004)]{GallimoreBeswick} 
Gallimore, J.~F., \& Beswick, R.\ 2004, \aj, 127, 239

\bibitem[Gebhardt et al.(2000)]{Gebhardt} Gebhardt, K., et al.\ 
2000, \apjl, 539, L13 

\bibitem[Gebhardt et al.(2011)]{Gebhardt2011} Gebhardt, K., Adams, 
J., Richstone, D., Lauer, T.~R., Faber, S.~M., G{\"u}ltekin, K., Murphy, 
J., \& Tremaine, S.\ 2011, \apj, 729, 119 

\bibitem[Genzel et al.(1998)]{Genzel} Genzel, R., et al.\ 
1998, \apj, 498, 579 

\bibitem[Gerssen et al.(2004)]{Gerssen} Gerssen, J., van der 
Marel, R.~P., Axon, D., Mihos, J.~C., Hernquist, L., 
\& Barnes, J.~E.\ 2004, \aj, 127, 75 

\bibitem[Ghez et al.(2008)]{Ghez} Ghez, A.~M., et al.\ 2008, 
\apj, 689, 1044

\bibitem[Gillessen et al.(2009)]{Gillessen} Gillessen, S., 
Eisenhauer, F., Trippe, S., Alexander, T., Genzel, R., Martins, F., 
\& Ott, T.\ 2009, \apj, 692, 1075 

\bibitem[Gonz{\'a}lez-Mart{\'{\i}}n et al.(2009)]{Gonz} 
Gonz{\'a}lez-Mart{\'{\i}}n, O., Masegosa, J., M{\'a}rquez, I., 
\& Guainazzi, M.\ 2009, \apj, 704, 1570

\bibitem[G{\"u}ltekin et al.(2009a)]{Gult_nuker} G{\"u}ltekin, K., 
et al.\ 2009a, \apj, 695, 1577

\bibitem[G{\"u}ltekin et al.(2009b)]{Gult} G{\"u}ltekin, K., 
et al.\ 2009b, \apj, 698, 198 

\bibitem[Hagiwara et al.(2011)]{Hagiwara} Hagiwara, Y., Baan, 
W.~A., \& Kl{\"o}ckner, H.-R.\ 2011, \aj, 142, 17 

\bibitem[Harms et al.(1994)]{Harms} Harms, R.~J., et al.\ 
1994, \apjl, 435, L35 

\bibitem[Herrnstein et al.(1999)]{maser} Herrnstein, J.~R., 
et al.\ 1999, \nat, 400, 539 

\bibitem[Hicks \& Malkan(2008)]{HicksMalkan} Hicks, E.~K.~S., 
\& Malkan, M.~A.\ 2008, \apjs, 174, 31 

\bibitem[Hopkins 
\& Quataert(2010a)]{HQ2010} Hopkins, P.~F., \& Quataert, E.\ 2010a, \mnras, 405, L41 

\bibitem[Hopkins 
\& Quataert(2010b)]{M31} Hopkins, P.~F., \& Quataert, E.\ 2010b, \mnras, 407, 1529 

\bibitem[Hopkins et al.(2006)]{Hopkins06} Hopkins, P.~F., 
Hernquist, L., Cox, T.~J., Di Matteo, T., Robertson, B., 
\& Springel, V.\ 2006, \apjs, 163, 1 

\bibitem[Kazantzidis et al.(2005)]{KazMayer} Kazantzidis, S., et 
al.\ 2005, \apjl, 623, L67 

\bibitem[Kiuchi et al.(2009)]{Kiuchi} Kiuchi, G., Ohta, K., 
\& Akiyama, M.\ 2009, \apj, 696, 1051 

\bibitem[Komossa et al.(2003)]{Komossa} Komossa, S., Burwitz, 
V., Hasinger, G., Predehl, P., Kaastra, J.~S., 
\& Ikebe, Y.\ 2003, \apjl, 582, L15 

\bibitem[Kormendy \& Gebhardt(2001)]{KormGeb} Kormendy, J., \& Gebhardt, K.\ 2001, 
in AIP Conf. Proc. Vol. 586, 20th Texas Symposium on Relativistic Astrophysics, ed. J.~C. Wheeler \& H. Martel (Melville, NY: AIP), 363 

\bibitem[Kormendy \& Richstone(1995)]{KormARAA} Kormendy, J., 
\& Richstone, D.\ 1995, \araa, 33, 581 

\bibitem[Larkin et al.(2006)]{Larkin06} Larkin, J., et al.\ 
2006, New Astronomy Review, 50, 362

\bibitem[Magorrian et al.(1998)]{Magorrian} Magorrian, J., et 
al.\ 1998, \aj, 115, 2285

\bibitem[Markwardt(2009)]{mpfit} Markwardt, C.~B.\ 2009, 
Astronomical Society of the Pacific Conference Series, 411, 251 

\bibitem[Max et al.(2007)]{Max07} Max, C.~E., Canalizo, G., 
\& de Vries, W.~H.\ 2007, Science, 316, 1877 

\bibitem[Max et al.(2005)]{Max05} Max, C.~E., Canalizo, G., 
Macintosh, B.~A., Raschke, L., Whysong, D., Antonucci, R., 
\& Schneider, G.\ 2005, \apj, 621, 738 

\bibitem[Mayer et al.(2007)]{Mayer} Mayer, L., Kazantzidis, 
S., Madau, P., Colpi, M., Quinn, T., 
\& Wadsley, J.\ 2007, Science, 316, 1874

\bibitem[McConnell et al.(2011)]{Nicholas} McConnell, N.~J., Ma, 
C.-P., Graham, J.~R., Gebhardt, K., Lauer, T.~R., Wright, S.~A., 
\& Richstone, D.~O.\ 2011, \apj, 728, 100 

\bibitem[McLean(2008)]{McLean} McLean, I.~S.\ 2008, Electronic 
Imaging in Astronomy: Detectors and Instrumentation (2nd ed.; New York, 
NY; Praxis Publishing, and Springer Science+Business Media)

\bibitem[Miyoshi et al.(1995)]{Miyoshi} Miyoshi, M., Moran, J., 
Herrnstein, J., Greenhill, L., Nakai, N., Diamond, P., 
\& Inoue, M.\ 1995, \nat, 373, 127 

\bibitem[Mor\'e(1978)]{lmfitting} Mor\'e, J.\ 1978, in Numerical Analysis, vol. 630, ed. G.~A.
Watson (Springer-Verlag: Berlin), 105

\bibitem[Neumayer et al.(2007)]{Neumayer} Neumayer, N., 
Cappellari, M., Reunanen, J., Rix, H.-W., van der Werf, P.~P., de Zeeuw, 
P.~T., \& Davies, R.~I.\ 2007, \apj, 671, 1329

\bibitem[Nowak et al.(2008)]{Nowak} Nowak, N., Saglia, R.~P., 
Thomas, J., Bender, R., Davies, R.~I., 
\& Gebhardt, K.\ 2008, \mnras, 391, 1629 

\bibitem[Oliva et al.(1999)]{OOMM99} Oliva, E., 
Origlia, L., Maiolino, R., \& Moorwood, A.~F.~M.\ 1999, \aap, 350, 9 

\bibitem[Onken et al.(2007)]{Onken07} Onken, C.~A., et al.\ 
2007, \apj, 670, 105 

\bibitem[Onken et al.(2004)]{Onken04} Onken, C.~A., Ferrarese, 
L., Merritt, D., Peterson, B.~M., Pogge, R.~W., Vestergaard, M., 
\& Wandel, A.\ 2004, \apj, 615, 645 

\bibitem[Pollack et al.(2007)]{Lindsay} Pollack, L.~K., Max, 
C.~E., \& Schneider, G.\ 2007, \apj, 660, 288

\bibitem[Riffel 
\& Storchi-Bergmann(2010)]{gasdisk} Riffel, R.~A., \& Storchi-Bergmann, T.\ 2011,
\mnras, 411, 469

\bibitem[Rothberg 
\& Fischer(2010)]{Rothberg} Rothberg, B., \& Fischer, J.\ 2010, \apj, 712, 318 

\bibitem[Sanders \& Mirabel(1996)]{SandersMirabel} Sanders, 
D.~B., \& Mirabel, I.~F.\ 1996, \araa, 34, 749 

\bibitem[Scoville et al.(2000)]{Scoville} Scoville, N.~Z., et 
al.\ 2000, \aj, 119, 991

\bibitem[Shields et al.(2003)]{Shields} Shields, G.~A., 
Gebhardt, K., Salviander, S., Wills, B.~J., Xie, B., Brotherton, M.~S., 
Yuan, J., \& Dietrich, M.\ 2003, \apj, 583, 124 

\bibitem[Siopis et al.(2009)]{Siopis} Siopis, C., et al.\ 
2009, \apj, 693, 946 

\bibitem[Tacconi et al.(1999)]{Tacconi} Tacconi, L.~J., Genzel, 
R., Tecza, M., Gallimore, J.~F., Downes, D., 
\& Scoville, N.~Z.\ 1999, \apj, 524, 732 

\bibitem[Tecza et al.(2000)]{Tecza} Tecza, M., Genzel, R., 
Tacconi, L.~J., Anders, S., Tacconi-Garman, L.~E., 
\& Thatte, N.\ 2000, \apj, 537, 178

\bibitem[Tremaine et al.(2002)]{Tremaine} Tremaine, S., et al.\ 
2002, \apj, 574, 740 

\bibitem[van Dam et al.(2006)]{vanDam06} van Dam, M.~A., et al.\ 
2006, \pasp, 118, 310 

\bibitem[Veilleux et al.(2009)]{Veilleux} Veilleux, S., et al.\ 
2009, \apj, 701, 587 

\bibitem[Wall \& Jenkins(2003)]{stats} Wall, J.~V., \& Jenkins, C.~R.\ 2003, Practical Statistics
for Astronomers, (Cambridge, UK; Cambridge University Press)

\bibitem[Winge et al.(2009)]{GNIRS} Winge, C., Riffel, R.~A., 
\& Storchi-Bergmann, T.\ 2009, \apjs, 185, 186

\bibitem[Wizinowich et al.(2006)]{Wiz06} Wizinowich, P.~L., 
et al.\ 2006, \pasp, 118, 297 


\end{thebibliography}
\end{document}